# Medical Image Computing and Computer Aided Medical Interventions applied to soft tissues.

# Work in progress in Urology.

[1]Jocelyne Troccaz and [1]M. Baumann, [1]P. Berkelman, [1]P. Cinquin, [1]V. Daanen, [1]A. Leroy, [1]M. Marchal, [1]Y. Payan, [1]E. Promayon, [1]S. Voros, [1,2]S. Bart, [3]M. Bolla, [2]E. Chartier-Kastler, [4]J.L.Descotes, [3]A. Dusserre, [3]J.Y. Giraud, [4]J.A. Long, [4]R. Moalic, [1,2]P. Mozer

[1]TIMC Laboratory

[2]Urology department, La Pitié Salpêtrière Hospital, Paris

[3]Radiotherapy Department, Grenoble Hospital

[4]Urology Department, Grenoble Hospital

**Abstract**

Until recently, Computer-Aided Medical Interventions (CAMI) and Medical Robotics have focused on rigid and non deformable anatomical structures. Nowadays, special attention is paid to soft tissues, raising complex issues due to their mobility and deformation. Mini-invasive digestive surgery was probably one of the first fields where soft tissues were handled through the development of simulators, tracking of anatomical structures and specific assistance robots. However, other clinical domains, for instance urology, are concerned. Indeed, laparoscopic surgery, new tumour destruction techniques (e.g. HIFU, radiofrequency, or cryoablation), increasingly early detection of cancer, and use of interventional and diagnostic imaging modalities, recently opened new challenges to the urologist and scientists involved in CAMI. This resulted in the last five years in a very significant increase of research and developments of computer-aided urology systems. In this paper, we propose a description of the main problems related to computer-aided diagnostic and therapy of soft tissues and give a survey of the different types of assistance offered to the urologist: robotization, image fusion, surgical navigation. Both research projects and operational industrial systems are discussed.

**Keywords**

Computer-aided surgery, medical robotics, medical image registration, urology.

## 1. Introduction

*1.1 A short introduction to urology*

Urology concerns the exploration, diagnostic and medical or surgical treatment of both the urinary apparatus of men and women and the genital apparatus of men. The organs of interest are the bladder, kidney, ureter, urethra and, for men, the prostate, penis and testicles (see Fig. 1). Pathologies include among others: lithiases (stones), cancers, traumas, stenoses, incontinence, infectious diseases, malformations and sterility. Urologic surgery also includes kidney transplantation. The major targets for robot or image-guided assistance are the prostate and the kidneys as detailed below.

Prostate cancer is one the most common malignancy among men. [Parkin01] reports year 2000 cancer statistics: 543000 cases and 204000 deaths were attributed to prostate cancer worldwide. Its detection is based on digital rectal examination (DRE) and Prostate Specific Antigen (PSA) rating and is confirmed through the anatomo-pathologic analysis of biopsies. Treatments include watchful waiting, surgery (laparoscopic or conventional radical prostatectomy), chemotherapy, and destruction of the tumour using different physical agents including radiotherapy (radiation by external beams), brachytherapy (radiation by implanted radioactive seeds), HIghly Focused Ultrasound, radiofrequency, and cryoablation. Because of the immediate anatomical environment of the prostate, in particular the bladder and rectum, and because of the role of this gland in the sexual life of patients, special attention is paid to minimally invasive techniques. One objective is to minimize induced morbidity. However, from the clinician standpoint the earlier detection of cancers from PSA screening and the development of laparoscopic techniques, by targeting smaller area via smaller entry ports, yield to increasing difficulties. Thus, computer or robot assistance may be needed.

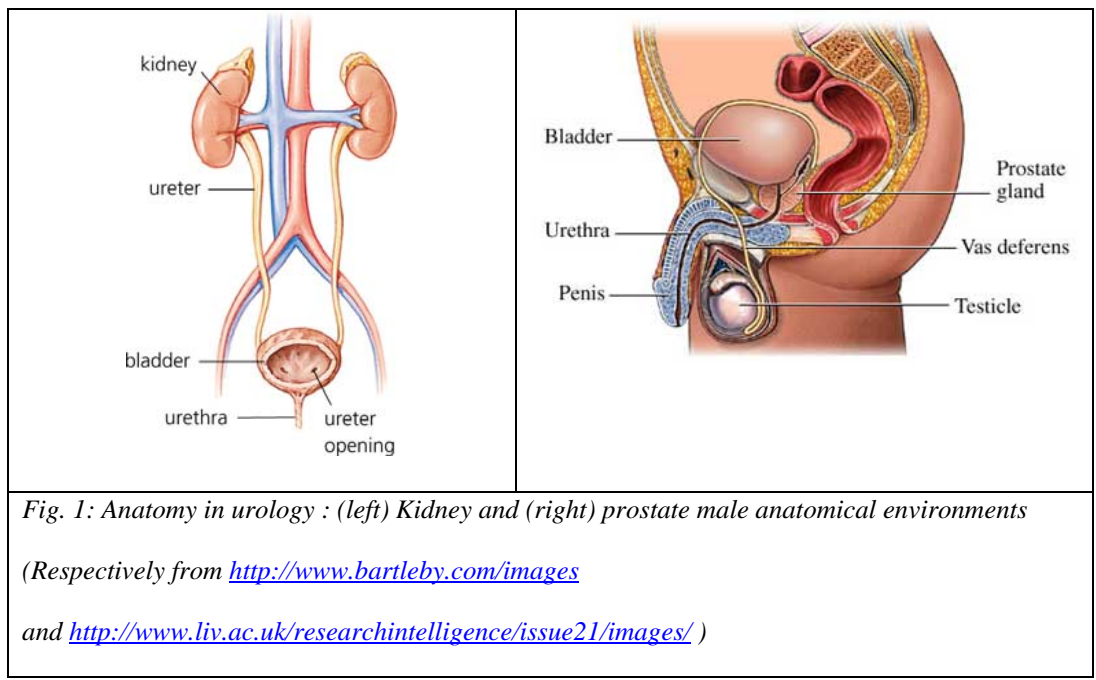

*Fig. 1: Anatomy in urology : (left) Kidney and (right) prostate male anatomical environments*

*(Respectively from http://www.bartleby.com/images*

*and http://www.liv.ac.uk/researchintelligence/issue21/images/ )*

Percutaneous access to the kidneys is also a challenging issue and concerns many patients. This technique can be used for any introduction of a needle in the kidney for diagnostic (biopsies) or therapeutic actions (radiofrequency cancer ablation or stone destruction for instance). The destruction of stones is a major clinical application: 5% of the occidental population is concerned. Traditionally, percutaneous access is controlled from real-time imaging (ultrasounds or fluoroscopy) whose drawbacks are respectively poor visibility and irradiation. There also, assistance would be welcome.

*1.2 Dealing with soft tissues*

Because urology deals with soft tissues, it is a perfect illustration of the difficulty to directly apply the computer-assistance know-how from bony structures to mobile and deformable tissues. Mobility and deformations have different origins:

- Intrinsic origin: some organs intrinsically move or are deformed to perform natural physiological activity such as breathing or cardiac rhythm. A foetus organ also has an intrinsic mobility due to foetal motion. Heart beating is quite predictable whilst foetal motion is not.
- Anatomical environment: other organs move or are deformed because of their anatomical environment. This is typically the case for the kidney that is moved up and down according to the diaphragmatic activity during breathing; this is also true for the prostate which position and orientation depend on the bladder and rectal filling, and in a lesser way, breathing.
- Patient position: the position and shape of some structures depend on the patient posture or position relatively to gravity. For example, the prostate position partly depends on the flexion of patient's legs.
- External action: finally, the therapeutic (needle insertion for instance) or diagnostic (e.g. ultrasound examination) action may move and deform the organ of interest. This is the case for the kidney and even more for the prostate, especially when an endorectal ultrasonic exam is performed.

Very often, the motion and deformation of an organ has multiple sources. For instance, the prostate moves and/or is deformed due to: patient breathing, patient posture, bladder and rectum natural or artificial filling, insertion of a needle, oedema from multiple needle insertions.

In the case of the prostate, several groups worldwide paid special attention in the mid-nineties to motion of the gland in the context of radiotherapy; since most intra-treatment localization approaches were based on X-Ray data where the prostate is not directly visible, it was important to quantify prostate motion with respect to bony structures. [VanHerk95] performed separate CT/MR bone and prostate registrations to determine prostate mobility; rigid chamfer matching on segmented surfaces was used; [Balter95] used implanted prostate fiducials and X-ray images to perform a similar study. More recently deformation was studied specially in the context of imaging involving intrarectal probes or coils (see for instance [Hirose02]).

In order to be able to handle these anatomic changes several issues must be solved: *models* must be designed when the motion and/or deformation are predictable and repeatable; *tracking* capabilities must be developed based on intra-operative sensing (images, signals such as ECG); *real-time re-planning* may be necessary for the guiding system (for instance a robot) to adapt to these changes; finally robots should be *synchronized* to those motions and deformations in a discrete or continuous way. This raises very challenging robustness and safety issues. One important characteristic of those anatomic changes is their time scale with respect to the duration of the action to be performed. Consequently, different strategies may be selected: localization just before the action or tracking during the whole action. In the following sections, the way those questions have been solved in the case of urological targets will be analyzed.

## 2. Robotics and urology

Historically, urology was one of the first clinical domains where a robot was used for patients. At the time – the late eighties – where most people dealt with neurosurgery or orthopaedics applications of robotics, the London Clinic and the Imperial College of London developed

PROBOT [Davies91]: a robot for the transurethral resection of the adenomatous prostate, i.e. the removal from the inside of the gland of extra-tissues compressing the urethra. The first test on a patient started in April 1991. After a feasibility study on 5 patients, a pre-clinical series with 40 patients was undertaken. Several versions of this system where developed; the first prototype was based on a PUMA 560 (from Unimation Inc.) connected to a passive frame. This frame is an elegant solution to safety issues since it constrains the tool movement inside a cone related to the task to be executed. The current system consists of a passive robot positioning a motorized frame with 3 degrees of freedom (dof) – conical motion plus translation of the resectoscope. [Shah01] reports the difficult task of automatically controlling this robot for resection monitoring from the real-time intra-operative ultrasound images.

Indeed, because soft tissues move and deform, two types of strategies may be used in robot control. The ideal approach would be to continuously and automatically close the robot control loop using intra-operative information about the organ motion. To our knowledge, such a solution has not yet been developed for urology. However in radiotherapy, where the tool is outside the body and the planning is rather simple (beam orientation with respect to. the patient and duration of radiation), organ tracking ability was introduced. In [Coste05] the motions of intra-body implanted fiducials are correlated to the motions of infra-red on-body markers for tracking breathing movements this process is however rather invasive. [Sawada04] proposes a non invasive solution based on real-time image correlation for the detection of a pre-defined stage in the breathing cycle (full expiration for instance); this information is used for respiratory-gated radiotherapy treatment. The other and much simpler approach is to tele-operate robots: in that case the user closes the loop between robot motion and real-time image information. Such an approach is particularly interesting when operative planning is too complex to be explicitly defined. Intermediate solutions consist in adding motion tracking abilities to tele-operated robots (see [Ginhoux05]) or to close the loop from imaging data in a more discrete way for simple tasks (see section 2.2).

*2.1 Tele-operated robots*

2.1.1 Endoscope holders

The first FDA[1] approved medical robot, AESOP (from Computer Motion Inc.) [Sackier94] had a significant clinical and industrial success. Two thousand AESOP were sold to around five hundred hospitals between years 1994 and 2000. AESOP has a SCARA architecture with 4 active and 2 passive (pivot rotation) dof; this tele-manipulator is voice controlled. Many other robotic endoscope holders have been developed in the academic and industrial tracks. One of them designed at TIMC [Berkelman03] has the interesting property of being directly put on the patient abdomen skin (see Fig. 2). Because the robot is placed on the endoscope entry point, 3 dof (2 rotations and 1 translation) are sufficient to handle the endoscope motions.

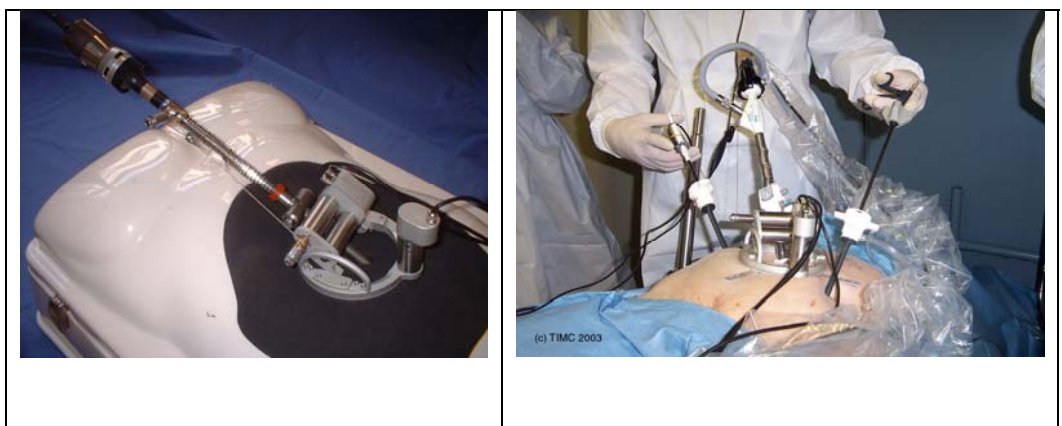

*Fig. 2.*

*The LER (Light Endoscopic Robot, TIMC, Grenoble Hospital and School of Medicine): on a phantom (left), urological intervention on cadaver (right).*

---

[1] Food and Drug Administration

As compared to AESOP and to most of the other systems which are positioned on the operating room (OR) table, floor or ceiling, this very compact system follows the patient motions and is very easy to install. It weights 625g; it is voice controlled and completely sterilizable. Interesting evolutions of robotic endoscope holders deal with automatically control of robots from image information in order to track organs or instruments during the surgery (see [Voros06] for instance).

2.1.2 Tele-surgery robots

Based on the robotic endoscope holders experience, instrument holders have naturally been designed resulting in the so-called tele-surgery robots. ZEUS, an evolution of the AESOP, is composed of 3 separated 4 dof arms (one endoscope holder and two instrument holders). Another system, the DaVinci (from Intuitive Surgical Inc.), is composed of 3 or 4 arms mounted on a single basis. Articulated instruments provide extra intra-body dof (see Fig. 3). Both systems are based on master-slave architectures; the arms are tele-operated[2] by the surgeon from endoscopic images. DaVinci proposes a "head-in" stereoscopic display (see Fig. 3) whilst Zeus includes a "head-mounted" stereoscopic display or a traditional screen. Intra-body dof are a major advantage of the DaVinci, increasing the surgeon's possibilities near open surgery conditions. Both systems are quite cumbersome and expensive; none of them include force feedback on the master workstation which may be a serious limitation for anastomoses for instance. The DaVinci has been extensively evaluated for urological applications. First robot-assisted laparoscopic radical prostatectomies were reported in [Abbou00], [Binder01]. Very large series of patients have since been operated: the Vattikuti Institute in the Henry Ford Hospital of Detroit, USA, published in [Menon04] a study concerning more than 1100 cases. In this centre, laparoscopic prostatectomies started in

---

[2]Technically, nothing really constrains the surgeon on the master console to be close to the slave robot; in practice – except for some concept demonstrations such as [Marescaux01] – the needs for reliability in data transmission and safety result in short-distance tele-surgery.

October 2000 and the DaVinci assistance was introduced in March 2001. A study comparing conventional/laparoscopic/robot-assisted laparoscopic procedures showed clear advantages of the robotic series on many points including shorter hospital stays, reduced pain, reduced blood loss, better PSA control, reduced positive margins, better continence, and less impotence. Another advantage of robot assistance is a reduction of the learning curve for laparoscopic procedures; [Alhering03] reports an improvement factor of about 10.

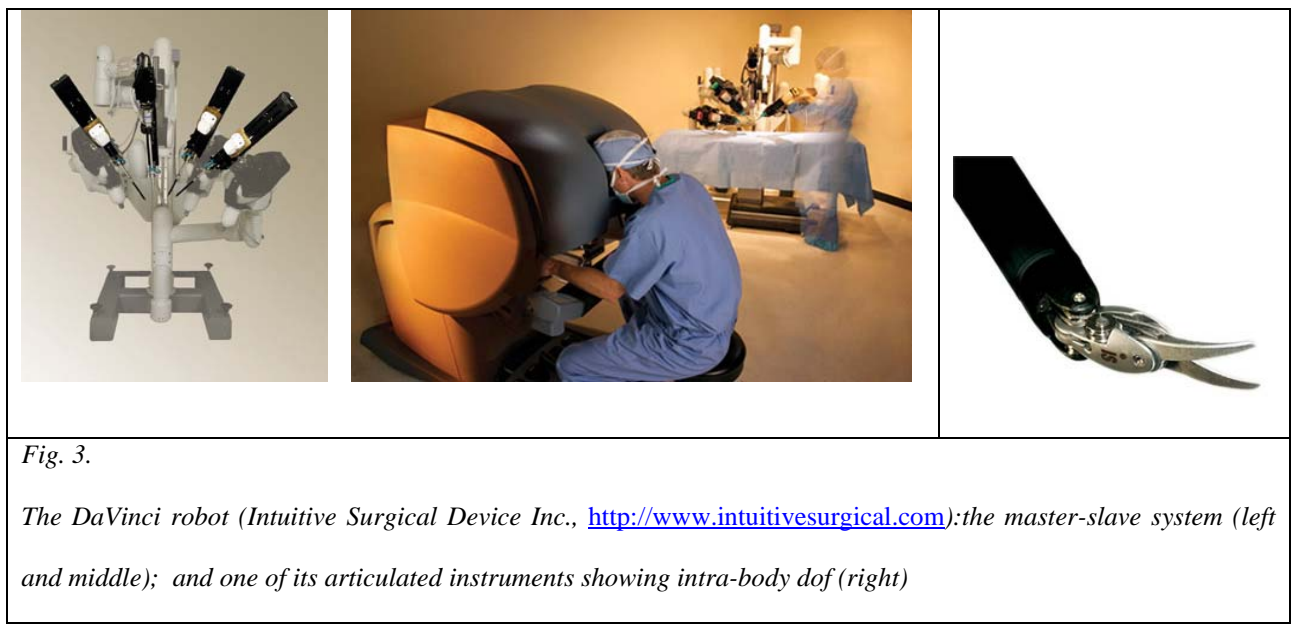

*Fig. 3.*

*The DaVinci robot (Intuitive Surgical Device Inc., http://www.intuitivesurgical.com):the master-slave system (left and middle); and one of its articulated instruments showing intra-body dof (right)*

Laparoscopic radical prostatectomy is probably one of the domains were the robotic clinical added-value was so clearly demonstrated. Other applications of such robots to urology are reported in full details in [UroCLin04]. Each time complex dissections, microsurgery or intra-corporal suturing are necessary, the robot may be a precious assistant.

Several research projects in the world aim at developing competitive smaller and/or cheaper solutions with articulated intra-body instruments and endoscopes and master station offering force feedback. Planning tools are also developed in order to optimize the entry ports positioning, enabling both target access and collision-free motion of the robots (see for instance [Coste04]).

*2.2 Image-guided robots*

Many gestures in urology are carried under interventional radiology: the diagnostic or therapeutic tool is moved under control of an imaging modality. Ultrasounds or fluoroscopy enable continuous control: the operator can see in real-time the tool position and the anatomy; CT or MRI allow asynchronous control: for instance, a needle is positioned, a control image is taken and the needle position is corrected if necessary, and so on. This idea has been exploited to control from medical images robots performing simple tasks such as a linear tool insertion.

2.2.1 Prostate biopsies and brachytherapies

From a technical viewpoint prostate biopsies and brachytherapies (see figure 4) are rather similar; they both consist in inserting needles in the prostate, either for tissue sampling or for radioactive seed placement, through transperineal or transrectal access, under imaging control – most often TransRectal UltraSound imaging (TRUS).

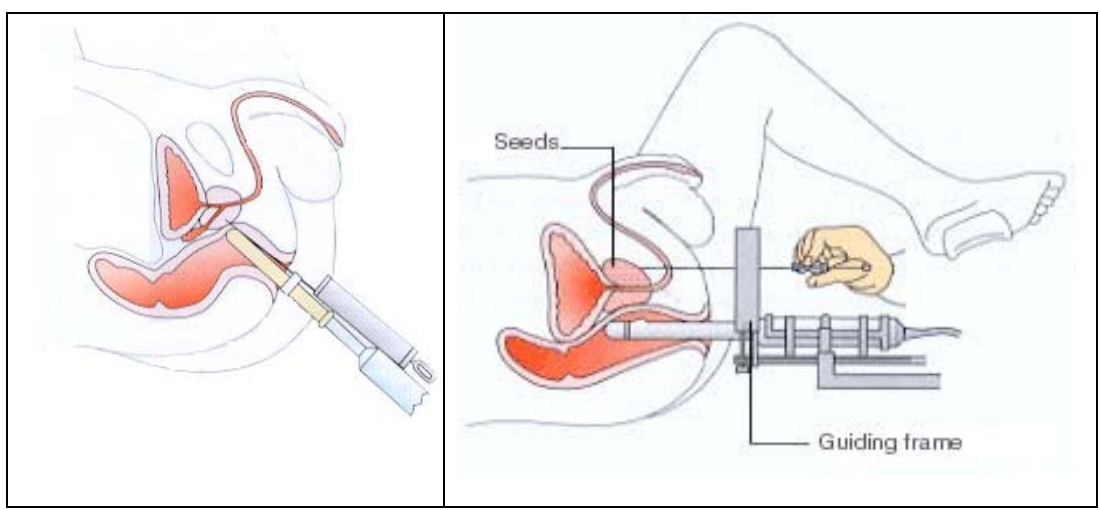

*Fig. 4.*

*US-guided transrectal biopsy (left) and transperineal brachytherapy (right) operating principles.*

*From* http://www.uropage.com/index.htm

However, each biopsy makes use of a single ultrasonic (US) image in which the needle is visible whilst brachytherapy is based on a volume of images: often parallel axial US images acquired every 5mm. Brachytherapy is based on a careful patient-specific dose planning whilst biopsies are generally performed following a predefined global scheme (for instance sextant or 11-core protocols). Needle insertion is slow and manual during brachytherapies whilst biopsies are very rapidly performed using a biospy gun. As demonstrated by [Heverly05] in a different medical context, increasing needle velocity results in minimizing the displacement and deformation of the tissue. Thus automation may have a positive impact in terms of gesture accuracy. Moreover, prostate brachytherapy is based on the use of a template (a stereotactic grid) rigidly connected to the US probe. This restrains needle trajectories to lines parallel to the probe axis and results in potential collisions of the needles with the pelvic bone. Again, using a robot may enable various trajectory directions. [Wei04] proposes to use a general purpose 6 dof robot for needle positioning and insertion. [Davies04] develops a special-purpose robot mimicking the conventional procedure (trajectories parallel to the US probe axis); a rotational dof for the needle is added to reduce the needle flexion during tissue penetration. Those systems are still laboratory test beds. [Phee05] describes a pre-clinical evaluation of a specific 9 dof system (positioning platform plus biopsy robot) for transperineal prostate biopsies. A 3D prostate geometric model of the prostate is approximated from series of close parallel US images enabling planning of the biopsies. 2.5mm accuracy is reported; those performances require very careful patient preparation and US probe handling. None of these systems really considers prostate motion and deformation during the procedure.

Another approach consists in performing transrectal prostate biopsies or brachytherapies with an intra-rectal robot under MRI control [Susil03]. Although conventional MR imaging (1.5T with endorectal antenna and T2 sequence) enables physicians to see precisely the prostate anatomy, using such a modality for biopsies is probably restricted to the few cases where US-guided biopsies are not possible or not successful. Let us remind that in the United States (resp. in France) about $10^6$ (resp. $10^5$) series of diagnostic biopsies are performed each year). In [Susil03], conventional MRI is used. The robot is inserted in the patient rectum and has three dof to reach the target defined on the MRI data: translation in the rectum, rotation around its main axis and progression of the needle. Thanks to its design, the robot does not disturb the magnetic field and includes two coils; one used as part of the imaging sensor and the other one as a position sensor. After validation on dogs, the robot is being clinically tested for transrectal biopsies and brachytherapies [Fichtinger04]; 2mm accuracy is reported; this remaining error is probably mainly due to the prostate motion and deformation during needle insertion. [Chinzei00] proposes another robot for prostate biopsy or brachytherapy inside an open interventional MR system. Interventional MRI (0.5T) requires an additional conventional MRI exam which makes the procedure even more complex (see 3.2.1).

2.2.2 Percutaneous renal access

The purpose is to assist percutaneous access to the kidney. Since 1996 a robot named PAKY (Percutaneous Access of the KidneY) is developed by the Johns Hopkins groups in Baltimore (MD, USA). The robot has seven passive dof used to position a 3 active dof structure (2 for orientation and 1 for translation of the needle). Fluoroscopy is used for needle alignment and control during insertion. During the procedure, the patient is in apnoea in order to keep the kidney in a constant position. [Cadeddu98] reports in vitro and in vivo experiments. In

[Su02], for 23 patients, no significant difference is reported between the manual and robotic procedure in terms of precision, rapidity, number of attempts, complications. One advantage of the robotic procedure lies in the absence of irradiation of the human operator. [Bascle00] proposes a visual servoing approach from two fluoroscopic views enabling the automatic placement of the needle to a given target and entry point. This system was also applied to CT-guided transperineal prostate biopsies through a single entry point.

**3. Image-based urology**

*3.1 Image processing*

Many papers propose tools to assist the segmentation of urologic images especially for the prostate where TRUS images have been paid close attention. Segmentation may be 2D, 2.5D (the segmentation of a given slice is used to help the segmentation of the following parallel one) or 3D. Most successful approaches make use of active contours and/or statistical models. However, for 2D images close to the prostate extremities, existing tools may not be robust enough due to the poor quality of those images. Other works concern the automatic segmentation of CT and MRI images of urological targets (kidney in particular). Because this problem is very vast and not typical of interventional systems, no details are given here. [Shao03] and [Zhu06] present good reviews of work concerning the image processing of prostate TRUS images.

*3.2 Image fusion*

MR and US imaging are probably the most used imaging modalities for prostate diagnosis and therapy. The interventional nature of US is counterbalanced by their traditional drawbacks: patient dependence, intra- and inter-operator variability, medium quality due to speckle, artefacts, etc. Conventional MRI using external or transrectal coils clearly show the prostate zonal anatomy which is useful for biopsy planning, whilst open MRI (also called iMRI for interventional MRI) enables near real-time control. This is why several research groups implemented fusion algorithm to benefit from complementary advantages of these modalities. Other imaging modalities are used such as CT imaging or histology sections; here also image fusion may be very useful. This paragraph describes different studies on multi-modality fusion dedicated to prostate imaging.

3.2.1 MRI/iMRI fusion

The Surgical Planning Laboratory (SPL) and Harvard Medical School have developed a navigation system for transperineal prostate biopsies under iMRI. Because of a lower intensity magnetic field, iMRI does not clearly show the prostate anatomy. This is why a pre-operative MRI acquisition is performed (external pubic antenna, 1.5T, T2 FSE sequence) on which surgical planning is possible. Intra-operatively, iMRI data are collected (external pubic antenna, 0.5T, T2 FSE sequence) enabling volume reconstruction. A localized stereotactic grid similar to the one used for brachytherapy is calibrated with respect to the MRI system and is used as a guiding device. MRI and iMRI are registered using an intensity-based elastic registration enabling transfer of the planning data to the iMRI conditions. During the biopsy procedure, real-time FGR iMRI slices (acquisition time: 8s per slice) are obtained for biopsy tracking. The "3D Slicer" developed by SPL enables computing from the iMRI T2-volume the slice corresponding to the FGR one with planning information added. [Hata01] reports two successful clinical cases. As mentioned previously, such an approach is especially interesting for patients for which ultrasounds have been unsuccessful (negative repeated US-guided biopsies with increasing PSA) or impossible. [Bharata01] describes another solution for MRI/iMRI registration based on elastic deformation of segmented data; a finite-element model of the pre-operative prostate is deformed using isotropic linear deformations to match the intra-operative prostate. Two regions of the prostate are considered with different elasticity parameters: the central gland and peripheral zone.

3.2.2 MRI/TRUS fusion

In many cases, bringing the MRI information to the US-guided procedures can be an interesting alternative. [Kaplan02] describes experiments concerning the fusion of pre-operative MRI data to TRUS images for transperinal biopsies. Six fiducial points are matched between the two modalities using a rigid registration. From this transform, a composite image is produced combining TRUS data and re-sliced matched MRI data. No attention is paid to motion or deformation that could occur between the two acquisitions or during the procedure.

In one of the two main types of brachytherapy protocols the radiation plan is prepared in the OR and is based on intra-operative TRUS acquisition and segmentation. One difficulty lies in this initial stage. [Reynier04] reports the evaluation of elastic surface registration between TRUS and MRI data for prostate brachytherapy. Pre-operatively, three orthogonal T2 TSE volumes are acquired using an endorectal coil and the prostate contours are segmented jointly on the three volumes. Intra-operatively, TRUS data are collected (axial and pseudo-sagittal slices – see Fig. 5-left). TRUS images are manually segmented by the urologist as in the conventional procedure and this results in a sparse intra-operative 3D prostate model (Fig. 5-middle); the axial contours are needed for simulating and planning the dose. A pre-registration consisting in superimposing the ultrasound and MRI data centres of gravity initializes the unknown transform before a second step of minimization using the Levenberg-Marquardt algorithm. Elastic registration allows for rotation and translation between data sets as well as local deformations or distortions. The method is derived from the octree-spline elastic registration described in [Szeliski96]. It makes use of an adaptive, hierarchical and regularized free-form deformation of one volume to the other coordinate system. The result is a 3-D function transforming any ultrasound data point to the corresponding MRI point. After MRI/TRUS surface-based elastic registration, composite MRI/TRUS images are computed; the operator can then visualize combined TRUS and MRI data of a same region (cf. Fig. 5-right). This enables refining the TRUS segmentations of the axial slices especially near the prostate extremities (base and apex). Elastic registration accounts for prostate motion and deformation between the two acquisitions. Processing of large intra-operative modifications due to needle insertions is limited to manual re-planning made by the radiophysicist during the procedure. [Reynier04] reports millimetre registration accuracy and shows that such a modified segmentation may result in a significantly different radiation plan.

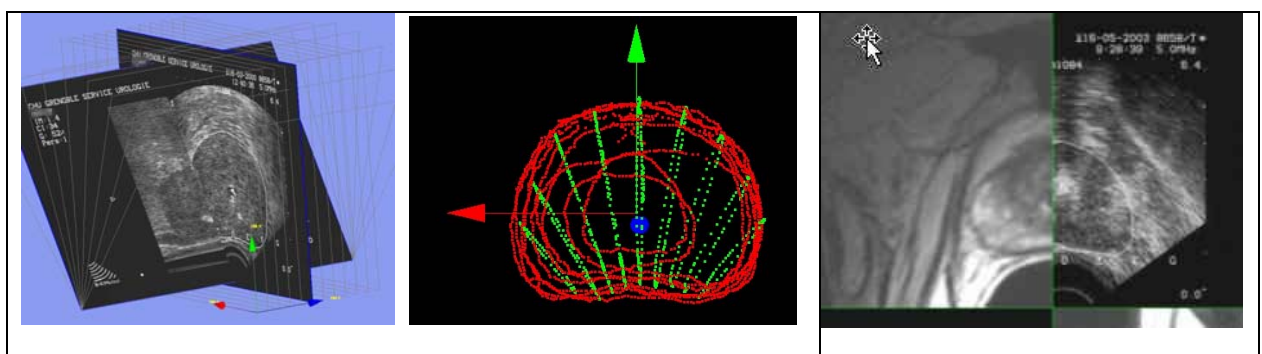

*Fig. 5.*

*IRM/TRUS fusion for brachytherapy (ProCur software, TIMC and Grenoble Hospital): (left) two of the axial and pseudo-sagittal TRUS images; (middle) 3D TRUS reconstruction; (right) composite image generated after MRI/TRUS elastic registration: for each axial TRUS image the corresponding MRI data are computed .*

3.2.3 CT/MR fusion

For radiation therapy (external radiation therapy or brachytherapy), CT imaging plays a central role in treatment planning or post-operative dose evaluation due to the need for tissue density information. Meanwhile, in prostate treatments, MR is still the main diagnostic modality. [Amdur99] presents CT/MR data fusion based on landmarks for seed implant quality assessment; the seeds are visible on post-operative CT images whilst the prostate is visible on pre-operative MR data. Rigid registration of CT and MR data is based on the use of a urinary catheter going through the prostate and placed before both MR and CT acquisitions making it visible on both modalities. After registration of the catheter, seeds visible on the CT data may be visualized with respect to the MR prostate anatomy. The authors report an average residual error measured on each slice at the centre of the urethra of 1.2mm for 11 patients and a maximum residual error of 2.5mm for one patient. [Lian04] proposes to use a Thin Plate Spline (TPS) function to map segmented prostate contours from MRI or MRSI (Magnetic Resonance Spectroscopic Imaging) to CT data. In this study, endorectal coils are used for MR imaging which may deform the prostate as compared to the CT acquisition situation; rigid registration may thus not be accurate enough. Obtained correspondence index of 93.1±5% and centroid position of 0.56±0.09mm of the registered surface slices quantify the registration accuracy and demonstrate the superiority of TPS registration over rigid one.

3.2.4 Histology/other modality fusion

Histology is the gold standard of cancer diagnosis. Therefore, registering histological data to other modalities may be very fruitful. After preparation and fixation in formalin of cancerous prostates surgically removed from patients, 3 to 5mm blocks are generally cut in a pseudo-axial direction; then a slice of 30 (to 50) microns thickness is obtained from each block and analyzed by a pathologist with a microscope for detection and characterization of cancerous cells. As proposed in [Egevad98] artificial markers are generally used in order to build a 3D reconstruction from those sparse data.

[Shen01] has fused histological data from hundreds of patients using elastic surface registration in order to build a statistical atlas of cancer occurrence in the different prostate

zones and to optimize biopsy schemes based on that atlas. [Bart05] has proposed to fuse MRI to histological data of patients with an elastic surface registration as a tool to study and improve the understanding of MR imaging of cancerous zones (see Fig. 6). [Taylor04] correlates histological data to US images and sonoelastometry. A first scaling factor is applied to account for prostate retraction during fixation; then, a rigid surface registration is applied. The purpose is here to evaluate the predictive value of sonoelastometry for tumour detection.

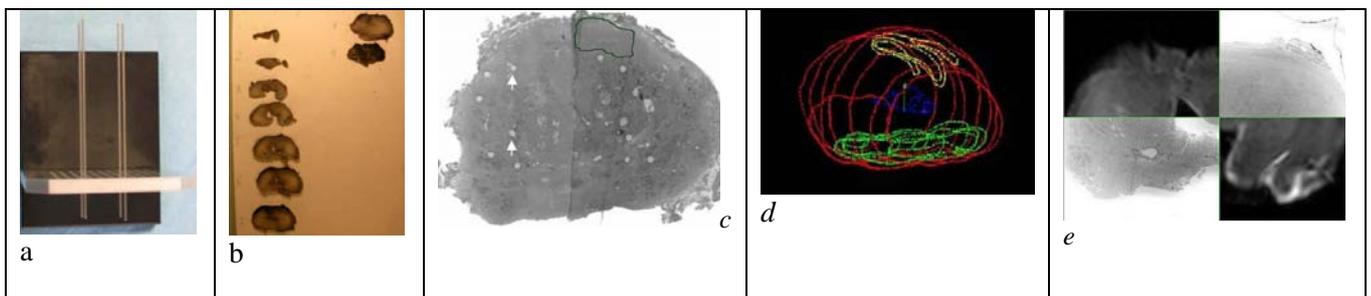

*Fig. 6.*

*Histology/MRI fusion (ProPath software, TIMC and Pitié Salpétrière Hospital, Paris): Needles and their guide for landmark definition inside the prostate (a); macroscopic 3mm slices of the resected prostate (b); slice reconstruction (arrows highlight some landmarks) (c); volume reconstruction including capsule, peripheral zone, urethra and cancer segmentations (d); composite image after MRI/histology elastic registration (e).*

*3.3 Surgical navigation*

Surgical navigation was introduced in neurosurgery in the eighties. It consists in tracking the surgical instruments relatively to poorly visible organs and/or to a pre-operative planning and to present this information visually to the surgeon. Those systems generally require trackers (optical, magnetic, US, mechanical) and involve data fusion and/or registering to enable transfer of the planning to the intra-operative conditions. Few urological applications exist.

3.3.1   Renal punctures

[Leroy04a,b, Mozer05] propose a system for navigating percutaneous access to the kidney (see Fig. 7). Pre-operatively two injected CT (early injection and late injection) are acquired under apnoea. Early injection allows visualization of the kidney envelope whilst late injection makes the internal structures of the kidney visible. Mono-modal rigid registration based on a chamfer matching is performed using the Analyze® software (from AnalyzeDirect Inc.). Planning is based on those registered data.

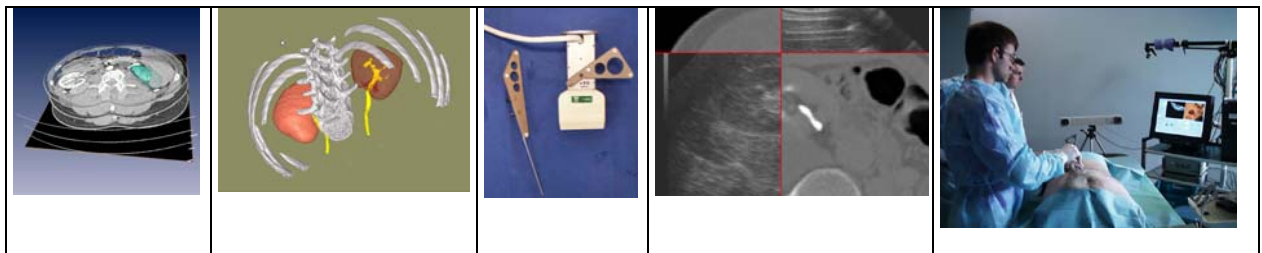

*Fig. 7.*

*Navigated percutaneous access of the kidney (TIMC, Pitié-Salpétrière Hospital, Grenoble School of Medicine, PRAXIM): (from left to right) scanner volume et segmented kidney; pre-operative planning interface; US probe and puncture needle with their optical localizers (passive Polaris from NDI, address) ; CT/US fusion; puncture guidance.*

Intra-operatively a localized ultrasonic probe enables pseudo-3D echographic acquisition of the kidney under apnoea; the patient is anesthetized. After a first version using surface-based registration and requiring segmentation of the data – CT was segmented with the 3D watershed of NablaVision software (from GenericVision, ApS) – an intensity-based rigid registration was implemented and tested. The similarity measurement is the correlation ratio. After registration, a graphical user interface guides the surgeon towards the planned trajectory and target position; the patient is again under apnoea. The system requires to the patient to be in the same apnoea configuration (full expiration for instance) during US acquisition and tool guidance. Experiments on volunteers have demonstrated a good repeatability of the kidney positioning under repeated apnoeas; this confirms literature results. Validation of the developed system on real data and on cadavers is promising. Real-time tracking of the kidney, for instance using magnetic localization, by avoiding the need for apnoea would greatly simplify the protocol; nevertheless the visual guidance may not be easy with a mobile target. Robotic alternatives may thus have to be considered for instance by synchronizing the robot to a given stage of the respiratory cycle. The motion of the kidney due to the needle insertion has not yet been studied. Here again, robotic access may minimize induced motion.

The same framework was applied to the guidance of stimulation electrode placement into the sacrum (S3 root) for incontinence treatment. CT/US surface-based registration of the pelvis originally developed for pelvic surgery (see [Tonetti01, Daanen04]) was successfully applied to this application and validated though cadaver experiments [Leroy04b]. In this application the bone position is tracked thanks to an implanted rigid body.

3.3.2 Prostate biopsies

The precise realization of prostate biopsy schemes (for instance twelve biopsies regularly distributed on the prostate gland) faces the difficulty of using 2D images to guide a 3D action. The process thus strongly depends on the surgeon's ability to mentally integrate successive images and trajectories in a 3D space. Because US-guided biopsies only detect 75% of the cancers for the first series, assisting biopsies to guarantee that samples are correctly and regularly acquired in the targeted sites is an important objective. An exploratory study using a navigation system (see Fig. 8) was developed at TIMC. The US probe is localized using an optical system (Passive Polaris from Northern Digital Inc.) and the executed trajectories are recorded in a fixed reference system. Those recordings clearly demonstrate that the prostate moves very significantly with the US probe displacements; those displacements are necessary to orient the needle, which is rigidly attached to the probe, towards the targeted sites.

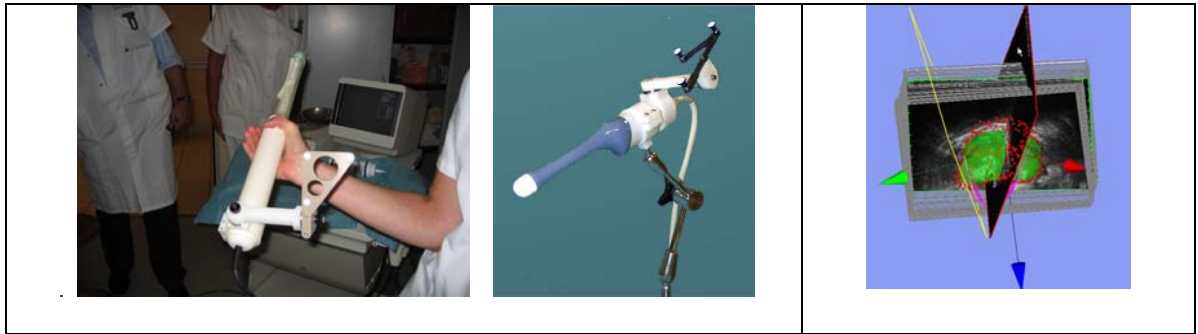

*Fig. 8.*

*Prostate biopsy navigation (ProNavV1, TIMC, Grenoble Hospital) – Biopsy recording: (left) localized intra-rectal US probes (2D and 3D) – (right) visualization interface presenting US images, prostate model and recorded trajectories in a fixed reference system.*

Current work deals with US-based real-time registration enabling to represent the executed biopsies in a single prostate reference frame despite prostate intra-operative displacements.

**4. Model-based urology**

As seen briefly in this paper, shape and/or appearance models have been introduced as a priori information for image processing. Statistical models concerning the occurrence of cancers in the different zones of the prostate have also been constructed in order to optimize the biopsy strategies by maximizing the ability to detect an existing tumour. [Schostak05] reports a clinical evaluation of such an optimized scheme.

In the context of medical interventions dealing with soft tissues much attention is paid to biomechanical models; the purpose is in particular to better predict tissue motions and deformations and tool interaction with the tissues. In a first stage these models were mostly developed for training simulators especially in laparoscopic surgery. More recently, they are seen as a necessary input to the planning of a surgery performed on soft tissues.

Some groups work on discrete models of the prostate complex environment; for instance [Kuroda03] presents a simulator for rectal palpation of the prostate. The interactive nature of such models enabling intra-operative re-planning may be counterbalanced by difficult parameter identification. Because of their theoretical background and the ability to introduce rheological tissue parameters, finite-element models are the most commonly used. Based on DiMaio and Salcudean previous work (cf. [DiMaio03]), [Goksel04] proposes a brachytherapy simulator combining TRUS image generation to a needle tissue interaction model. [Mohammed02] combines statistical and biomechanical models for evaluation of intra-operative prostate deformation. Other groups work on steerable needle path planner taking benefit of needle flexion to generate paths avoiding anatomical obstacles [Alterovitz05].

Current limitations of planning approaches are due to the hypothesis of a homogeneous tissue and, for patient-specific planners, in the difficulty of determining in-vivo tissue physical parameters. Sonoelastometry, a very promising modality for the detection of cancers inside organs [Konig05], can also help for physical parameters identification.

**Conclusion**

As seen in this paper, many research projects are dedicated to urological applications. Concerning target motion and deformation, some partial solutions have been proposed. Tele-operated robots enable adapting the surgical strategy to the anatomical state based on the surgeon know-how. Image-guided robots generally allow data acquisition just before the action to be executed; however none of them, in urology, is completely controlled by real-time data. Image fusion can consider changes occurring between the different acquisitions stages to be registered; however, no system yet enables real-time fusion of intra-operative data to pre-operative data for tracking purpose. Regarding the kidney, navigational assistance or robot actions are performed on a stabilized organ (breathing is temporarily held); no tracking is available yet. Biomechanical models are developed but none of them is yet used for guiding the intervention.

A lot has been done but significant research efforts must still be undertaken to fully consider a mobile and deformable target such as the kidney or the prostate. Intra-operative tracking based on intra-body markers (for instance magnetic markers or other [Coste05, Bricault05]) or on real-time image processing is necessary. The development of autonomous robots servo-ed to real-time intra-operative data (see for instance [Vitrani05]) raises very challenging robustness and safety issues since in this case the surgeon will no longer be able to directly supervise the robot actions. Finally, patient-specific modelling of organs mechanical properties is a key issue for predicting and recognizing anatomical changes and to allow precise planning. All those topics open very interesting and difficult scientific tracks for the coming years with a very strong clinical interest.

**Acknowledgements**

Research projects of TIMC described in this paper are/were supported by grants from the French Ministry of Health (PHRC 2003 Prostate-Echo, CIT PICAMI), the French Ministry of Research and ANVAR (MMM project in the RNTL Program), non-profit organizations (Association pour la Recherche contre le Cancer, Association Nationale de la Recherche Technique) and by Praxim-Medivision.